\documentclass[10pt]{article}
\usepackage{latexsym}
\usepackage{epsfig}
\usepackage{color}
\usepackage{amsmath}

\newcommand{\lbl}[1]{\label{eq:#1}}
\newcommand{ \rf}[1]{(\ref{eq:#1})}
\newcommand{\vs}[1]{\rule[- #1 mm]{0mm}{#1 mm}}
\newskip\humongous \humongous=0pt plus 1000pt minus 1000pt

\newif\ifdtup

\newcommand{\be}{\vs{0}\begin{equation}}
\newcommand{\ee}{\\[0mm]\end{equation}}
\newcommand{\bea}{\begin{eqnarray}}
\newcommand{\eea}{\end{eqnarray}}
\newcommand{\lapprox}{%
\mathrel{%
\setbox0=\hbox{$<$}
\raise0.6ex\copy0\kern-\wd0
\lower0.65ex\hbox{$\sim$}
}}
\newcommand{\gapprox}{%
\mathrel{%
\setbox0=\hbox{$>$}
\raise0.6ex\copy0\kern-\wd0
\lower0.65ex\hbox{$\sim$}
}}

\begin{document}
\allowdisplaybreaks

\indent

\begin{center}
{\bf\Large{Predictions for the rare kaon decays $K_{S,L} \to \pi^0 \ell^+ \ell^-$ from QCD\\[0.15cm] 
in the limit of a large number of colours}}

\indent

{\bf  
Giancarlo D'Ambrosio$^1$ and Marc Knecht$^2$
}
\\

\textit{$^1$INFN-Sezione di Napoli, Complesso Universitario di Monte S. Angelo,
Via Cintia Edificio 6,}\\
\textit{80126 Napoli, Italy}

\textit{$^2$Centre de Physique Th\'eorique, Aix-Marseille Univ./Univ. de Toulon/CNRS (UMR 7332),\\
CNRS-Luminy Case 907, 13288 Marseille Cedex 9, France}

\end{center}

\begin{abstract}
\noindent
The long-distance and non-local parts of the form factors describing the single-photon mediated
$K_{S,L}\to\pi^0\gamma^*\to\pi^0\ell^+\ell^-$ ($\ell = e, \mu$) transitions in the standard model
are addressed in QCD in the limit where the number $N_c$ of colours becomes infinite. It is 
shown that this provides a suitable theoretical framework to study these decay modes and 
that it allows to predict the decay rates for $K_{S}\to\pi^0\ell^+\ell^-$. 
It also unambiguously predicts that the interference between the direct and indirect
CP-violating contributions to the decay rate for $K_L\to\pi^0 \ell^+ \ell^-$ is constructive.
\end{abstract}

\twocolumn

Rare kaon decays remain a very active domain of research, with quite interesting
perspectives for the future, as attested by several recent reports \cite{AlvesJunior:2018ldo} - \cite{Nanjo:2023xvj}.
Since they are mediated by neutral currents, these processes are naturally suppressed in the standard model
\cite{Glashow:1970gm} \cite{Glashow:1976nt} and provide various ways to test the standard-model's flavour structure.
A fruitful completion of this research program requires a high 
level of precision in both experimental measurements 
and theoretical predictions. This goal is about to be fulfilled on the theory side
\cite{Buras:2021nns} \cite{Brod:2021hsj} \cite{Aebischer:2022vky} in the case of the rare decay modes 
$K\to\pi\nu{\bar\nu}$, which are dominated by short-distance contributions, 
and the prospects to improve on present experimental results \cite{KOTO:2018dsc} \cite{NA62:2021zjw}
look also quite promising \cite{HIKE:2022qra} \cite{Nanjo:2023xvj}. Unfortunately, the situation
is in a less satisfactory state, at least from the theoretical point of view, in the case of other rare kaon decay modes,
whose amplitudes are instead dominated by a long-distance and non-local component that
is governed by the non-perturbative dynamics of the strong interactions (QCD)
at low energies.

In the present Letter we wish to address this issue in the case of the decay modes of the 
neutral kaons $K_S$ and $K_L$ into a neutral pion and a pair of charged leptons. In the case of 
the short-lived neutral kaon, we will consider the CP-conserving transition mediated by 
the exchange of a single virtual photon, $K_S \to \pi^0 \gamma^* \to \pi^0 \ell^+ \ell^-$, with $\ell = e, \mu$,
and $K_S$ being identified with the CP-even combination $K_1^0$ of $K^0$ and ${\overline K} { }^0$, i.e.,
using the convention ${\rm CP} \vert K^0 \rangle = - \vert{\overline K} { }^0\rangle$,
\be
\vert K_S \rangle \simeq \vert K_{1}^0 \rangle = \frac{\vert K^0 \rangle - \vert {\overline K} { }^0 \rangle }{\sqrt{2}}   .
\ee
In the case of the long-lived kaon, defined as
\be
\vert K_L \rangle \simeq \vert K_2^0 \rangle + {\bar\epsilon} \vert K_1^0 \rangle ,
~~ \vert K_{2}^0 \rangle = \frac{\vert K^0 \rangle + \vert {\overline K} { }^0 \rangle }{\sqrt{2}}   ,
\ee
the situation is so to say reversed: while conservation of CP requires the exchange of two 
virtual photons, $K_2^0 \to \pi^0 \gamma^* \gamma^* \to \pi^0 \ell^+ \ell^-$,
the transition $K_2^0 \to \pi^0 \gamma^* \to \pi^0 \ell^+ \ell^-$
corresponds to a direct violation of CP \cite{Donoghue:1987awa}. It has been argued \cite{Buchalla:2003sj} and is usually admitted  
\cite{Isidori:2004rb} \cite{Mescia:2006jd} that the 
corresponding contribution to the amplitude is dominated by short distances,
and is thus proportional, in the standard model, to ${\rm Im}\,\lambda_t>0$, with $\lambda_t \equiv V_{td} V_{ts}^*$
a product of CKM matrix elements \cite{Dib:1988md}. 
Finally, the amplitude for this process receives a third component, due to CP violation in the mixing, which results from the 
non-vanishing parameter ${\bar\epsilon}$. For the latter we will take \cite{Donoghue:1987awa} \cite{Buchalla:1996fp} \cite{ParticleDataGroup:2022pth}
\be
{\bar\epsilon} \sim \frac{1+i}{\sqrt{2}} \vert\epsilon\vert , ~ \vert\epsilon\vert = 2.228 \cdot 10^{-3} .
\ee
The branching ratio thus takes the form \cite{DAmbrosio:1998gur} \cite{Buchalla:2003sj} \cite{Isidori:2004rb} \cite{Mescia:2006jd}
\bea
{\rm Br} (K_L \to \pi^0 \ell^+ \ell^-) \!\!\!&=&\!\!\! 10^{-12}
\Bigg[ C_{\rm mix}^{(\ell)} + C_{\rm int}^{(\ell)} \frac{{\rm Im} \, \lambda_t}{10^{-4}}
\nonumber\\
&&\!\!\!
+ C_{\rm dir}^{(\ell)} \left( \frac{{\rm Im} \, \lambda_t}{10^{-4}} \right)^2 + C_{\gamma^*\gamma^*}^{(\ell)} \Bigg]   . ~~~~~
\lbl{Br_KL}
\eea
The last term in this expression is the CP-conserving component. Phenomenological
estimates have found that it is small in the case 
$\ell=e$, $C_{\gamma^*\gamma^*}^{(e)} = {\cal O} (10^{-2})$ \cite{Donoghue:1987awa} \cite{Ecker:1987hd} \cite{Buchalla:2003sj}, 
and substantial in the case $\ell=\mu$,
$C_{\gamma^*\gamma^*}^{(\mu)} = 5.2(1.6)$ \cite{Isidori:2004rb} \cite{Mescia:2006jd}.
The first term in eq. \rf{Br_KL} gives the contribution from indirect CP violation alone,
and can be expressed in terms of experimental quantities \cite{Buchalla:2003sj},
the lifetimes $\tau(K_{S,L})$ of the neutral kaons, and the branching ratio
for the CP-conserving transition $K_S \to \pi^0 \ell^+ \ell^-$,
\be
C_{\rm mix}^{(\ell)} = 10^{12} \vert{\bar\epsilon}\vert^2 \frac{\tau(K_S)}{\tau(K_L)} {\rm Br} (K_S \to \pi^0 \ell^+ \ell^-)   .
\lbl{Cmix_exp}
\ee
The third term in eq. \rf{Br_KL} is the contribution from direct CP violation,
while the second term gives the interference between direct and indirect CP-violating contributions. Their dependence 
with respect to $\lambda_t$ is shown explicitly. The coefficient $C_{\rm int}^{(\ell)}$ is given as a 
phase-space integral whose integrand involves the amplitude of the decay $K_S \to \pi^0 \ell^+ \ell^-$.
A crucial issue is whether this interference is constructive or destructive: from an experimental point of view, 
a constructive interference will be a key feature in order to overcome the important irreducible background 
induced by the $K_L \to \gamma \gamma \ell^+ \ell^-$ decay \cite{Greenlee:1990qy}, and thus provide access to 
an independent determination of ${\rm Im} \, \lambda_t$. This brief description of the decays $K_{S,L} \to \pi^0 \ell^+ \ell^-$
leaves us with a short list of questions to be answered:
\begin{itemize}
 \item Can one predict ${\rm Br} (K_S \to \pi^0 \ell^+ \ell^-)$ (or even
 the decay distribution) in the standard model?
 \item Can the sign of $C_{\rm int}$ be predicted?
 \item Can one confirm that the long-distance component 
 of the amplitude induced by the direct-CP violating contribution $K_2^0 \to \pi^0 \gamma^* \to \pi^0 \ell^+ \ell^-$ remains 
 indeed negligeable once non-perturbative QCD effects are taken into account?
\end{itemize}
Answering these questions requires to obtain a quantitative control of 
the non-perturbative aspects of QCD at low energies, a notoriously difficult task.
The purpose of this Letter is to show that this goal can be met in the limit where 
the number of colours $N_c$ becomes infinite \cite{tHooft:1973alw} \cite{Witten:1979kh},
a limit which has often provided relevant insights into the 
physical case $N_c=3$. It turns out that in this large-$N_c$ limit QCD
leads to unambiguous positive answers for all the three questions listed above.
In order to show this, it is necessary that we first state more precisely in which manner 
the large-$N_c$ limit of QCD can be implemented in the case at hand.

Long-distance dominated rare kaon decays are traditionally addressed within 
the framework of the three-flavour low-energy expansion (ChPT) \cite{Gasser:1984gg}
extended to weak decays \cite{Cronin:1967jq} \cite{Kambor:1989tz} \cite{Esposito-Farese:1990inm} \cite{Ecker:1992de}. 
The lowest-order (one loop in this case) expression of the amplitudes
for $K\to\pi\ell^+\ell^-$, $(K,\pi)$ $=(K^\pm,\pi^\pm), (K_S , \pi^0)$ 
were obtained in ref. \cite{Ecker:1987qi} (see also \cite{Ananthanarayan:2012hu})
and expressed in terms of form factors ${\cal W}_+ (s)$ and ${\cal W}_S (s)$,
where $s$ denotes the square of the invariant mass of the di-lepton pair. 
A `beyond-one-loop' representation of these form factors, accounting only for part 
of the pion loops at next-to-lowest order, was proposed in ref. \cite{DAmbrosio:1998gur}
and reads
\be
{\cal W}_{+,S} (s) = G_F ( M_K^2  a_{+,S} + b_{+,S} s) + {\cal V}_{+,S}^{\pi\pi} (s)  . 
\lbl{rep_W_S}
\ee
The neglected contributions from pion loops were shown to be indeed
smallish in the whole range of energies corresponding to the 
relevant kinematic region \cite{DAmbrosio:2018ytt}. The counter-terms 
at lowest and at next-to-lowest orders as well as the loops involving  
also kaons, i.e. from $K{\overline K}$ intermediate states (already at 
one loop) or from $K\pi$ intermediate states (starting at two loops),
corresponding to higher thresholds sufficiently far away from the 
decay region, are described by a first-order polynomial in $s$. The expressions
for the contributions ${\cal V}_{+}^{\pi\pi} (s)$ and ${\cal V}_{S}^{\pi\pi} (s)$
from the pion loops are given in ref. \cite{DAmbrosio:1998gur}.
Focusing on $K_S \to \pi^0 \ell^+ \ell^-$, it turns out that ${\cal V}_{S}^{\pi\pi} (s)$
is suppressed, since it proceeds through a $\Delta I = 3/2$ transition 
$K_S \to \pi^0 \pi^+ \pi^-$. Predicting 
the decay distribution and decay rate therefore amounts, in practice, 
to being able to predict the values of the two unknown parameters $a_S$ and $b_S$.
Quantitative information about $a_S$ and $b_S$
is not provided by ChPT itself and needs to be looked for in 
the non-perturbative regime of full QCD. This is where we can expect 
that the limit of a large number of colours may become useful. Indeed, these two constants, 
or more precisely the contributions from the counter-terms to them, are precisely 
what survives from the amplitude \rf{rep_W_S} at leading order in the limit 
$N_c\to\infty$, since 
\be
a_{+,S} , b_{+,S} \sim {\cal O} (N_c) , ~~ {\cal V}_{+,S}^{\pi\pi} (s) \sim {\cal O} (N_c^0) .
\ee
Obtaining the representation of the 
form factor ${\cal W}_S (s)$ in the large-$N_c$ limit of QCD 
should therefore provide a good description of the amplitude 
in the decay region. In the remainder of this Letter
we will outline the main steps of this endeavour, relying 
partly on ref. \cite{MK_in_prep}, where a more detailed account
will be given, while here we merely discuss some phenomenological consequences. 
Before proceeding, let us mention that a similar procedure can 
be applied to the amplitude ${\cal W}_+ (s)$ as well, and we briefly comment on it 
before concluding this study. A more detailed discussion of ${\cal W}_+ (s)$ in the large-$N_c$ limit 
will be given in ref. \cite{MK_in_prep}.

In the standard model, the structure of the amplitude ${\cal A}_S$ of the decay
$K_S (k) \to \pi^0 (p) \ell^+ (p_+) \ell^- (p_-)$, with $k-p=p_+ + p_-$ and 
$s=(k-p)^2$, reads
\be
{\cal A}_S = {\cal A}^{\mbox{\tiny SD;A}}_S - e^2 
{\bar{\rm u}} (p_{\ell^-}\!) \gamma_\rho {\rm v} (p_{\ell^+}\!)
 (k + p)^\rho \times \frac{{\cal W}_S (s)}{16 \pi^2 M_K^2}   .
 \lbl{amp}
\ee
Let us for the moment leave aside the short-distance part ${\cal A}^{\mbox{\tiny SD;A}}_S$
and concentrate on the form factor ${\cal W}_{S} (s)$. It comprises another local
short-distance part, but also a long-distance dominated, non-local component,
\be
{\cal W}_{S} (s) = {\cal W}_{S}^{\mbox{\scriptsize loc}} (s ;\nu) + {\cal W}_{S}^{\mbox{\scriptsize non-loc}} (s ;\nu)   .
\lbl{FF_decomp}
\ee
The latter is given by
\bea
&&\hspace{-0.75cm}
\left[ s (k + p)_\rho - (M_K^2 - M_\pi^2) (k - p)_\rho \right]
\times \frac{W_{S}^{\mbox{\scriptsize non-loc}} (s ; \nu)}{16 \pi^2 M_K^2} 
\nonumber\\
&&
= i \! \int \! d^4 x \,
\langle\pi^0 (p)  \vert T \{  j_\rho  (0) {\cal L}^{\vert\Delta S\vert = 1}_{\mbox{\scriptsize non-lept}} (x)  \} \vert K_S(k) \rangle_{\mbox{\tiny$\overline{\rm MS}$}}  , ~~~~~
\lbl{FF_non-loc}
\eea
where $j_\rho$ denotes the three-flavour electromagnetic current,
\vspace{-0.5cm}
\be 
j_\rho (x) = \frac{2}{3} ({\bar u} \gamma_\rho u) (x) - \frac{1}{3} [ ({\bar d} \gamma_\rho d) (x) + ({\bar s} \gamma_\rho s) (x) ]  ,
\ee
and ${\cal L}^{\vert\Delta S\vert = 1}_{\rm non-lept}$ is the order ${\cal O} (G_{\rm F})$
effective Lagrangian for $\vert\Delta S\vert = 1$ weak non-leptonic transitions below the charm-quark 
threshold,
\be
{\cal L}^{\vert\Delta S\vert = 1}_{\rm non-lept}  = - \frac{G_{\rm F}}{\sqrt{2}} V_{us} V_{ud} 
\sum_{I=1}^{6}
C_I (\nu) Q_I (\nu) + {\rm H. c.}   
\ee
The current-current four-quark operators $Q_1$ and $Q_2$ read ($i$ and $j$ are colour indices)
\be
Q_1 = ( {\bar s}^i u_j )_{V-A} ( {\bar u}^j d_i )_{V-A}  , \,
Q_2 = ( {\bar s}^i u_i )_{V-A} ( {\bar u}^j d_j )_{V-A} .
\lbl{Q1_Q2}
\ee
The QCD-penguin operators $Q_{3,4,5,6}$ are given in e.g. ref. \cite{Buras:1994qa}.
In this same reference, the anomalous dimensions of these four-quark operators are also computed at next-to-leading order, 
which allows to evolve the Wilson coefficients $C_I(\nu)$ from the electroweak scale $\nu=M_W$, where they are computed to order
${\cal O} (\alpha_s)$, down to the low scale $\nu\gapprox 1~{\rm GeV}$, thus including, in a
renormalization-group improved perturbative way, with resummation of leading and next-to-leading
logarithms, all contributions generated by the degrees of freedom between $M_W$ and $\nu$. 
For $\nu$ below $1~{\rm GeV}$ this perturbative treatment can no longer be trusted, and the contributions from degrees of 
freedom below $1~{\rm GeV}$ are then provided by the non-perturbative matrix elements of the four-quark operators between hadronic 
states. While ChPT provides the contributions of the light pseudoscalar mesons to these matrix elements,
it cannot account fully, that is otherwise than by largely unknown counter-terms \cite{Ecker:1987qi} \cite{Ananthanarayan:2012hu}, 
for those of the hadronic resonances in the $1~{\rm GeV}$ region. As we will see,
this is where the large-$N_c$ limit steps in as an interesting alternative.

Notice that although ${\cal L}^{\vert\Delta S\vert = 1}_{\rm non-lept}$ does not depend on the separation scale $\nu$,
the decomposition \rf{FF_decomp} does depend on it. This is a consequence of the fact that 
the definition of the non-local part ${\cal W}_{S}^{\mbox{\scriptsize non-loc}}$ of the form factor involves a time-ordered product 
that is singular at short distances \cite{Isidori:2005tv} \cite{DAmbrosio:2018ytt}, for instance
(square brackets indicate colour-singlet quark bilinears.)
\bea
&&\hspace{-0.8cm}
\lim_{x\to 0}
 T \{ j_\rho (x) Q_1(0) \} \sim
\nonumber\\
&&\hspace{-0.75cm}
- \frac{N_c}{18 \pi^4}  
 [ {\bar s}  \gamma_\mu (1 - \gamma_5) d ] (0)
 \left( \delta_\rho^\mu \Box - \partial_\rho \partial^\mu \right) \frac{1}{(x^2)^2} + \cdots, ~~~~~
 \lbl{SD_sing}
\eea
where the ellipsis denotes subdominant corrections. This requires that
the time-ordered product in eq. \rf{FF_non-loc} be first regularized,
here we have used dimensional regularization, and eventually renormalized,
here in the ${\overline{\rm MS}}$ scheme, as indicated by the subscript on
the right-hand side of eq. \rf{FF_non-loc}, leaving behind a dependence with respect
to the associated renormalization scale $\nu$ in $W_{S}^{\mbox{\scriptsize non-loc}} (s ; \nu)$.
In this renormalization process the divergent part in the time-ordered product in
eq. \rf{FF_non-loc} has to be absorbed by a local counter-term. The latter is 
provided by the Gilman-Wise operator $Q_{7V}$ \cite{Gilman:1979bc}: for a complete description of the form factor
${\cal W}_S (s)$ in the standard model one also needs to consider contributions
from
\be
{\cal L}^{\vert\Delta S\vert = 1}_{\rm lept} (\nu) =
- \frac{G_{\rm F}}{\sqrt{2}} V_{us} V_{ud} \left[
C_{7V} (\nu)  Q_{7V}  + C_{7A} Q_A \right] + {\rm H. c.} 
\ee
involving two additional local four-fermion operators with a mixed quark$\times$lepton content,
\bea
&&
Q_{7V} = ({\bar\ell} \gamma_\mu \ell)[{\bar s} \gamma^\mu (1 - \gamma_5) d] ,
\nonumber\\
&&
Q_{7A}= ({\bar\ell} \gamma_\mu \gamma_5 \ell)[{\bar s} \gamma^\mu (1 - \gamma_5) d] .
\eea
These operators are both finite, and the scale dependence of the Wilson
coefficient $C_{7V} (\nu)$ can be interpreted as 
resulting from the absorption by a `bare coupling' $C_{7V}^{\mbox{\scriptsize bare}}$
of the local divergence of the time-ordered product in eq. \rf{FF_non-loc}.
The scale dependence has to cancel between the two contributions once they are added up
to form the physical form factor ${\cal W}_S (s)$ in eq. \rf{FF_decomp}.
A general discussion of how this happens, at least at next-to-leading
order in perturbative QCD, can be found in 
refs. \cite{DAmbrosio:2018ytt} \cite{DAmbrosio:2019xph}, and it
carries over to the limit $N_c\to\infty$ \cite{MK_in_prep}.
Finally, the operator $Q_{7A}$ and its Wilson coefficients $C_{7A}$
are defined at the electroweak scale, and need not be renormalized
in the standard model. This operator provides the contribution ${\cal A}^{\mbox{\tiny SD;A}}_S$
to the amplitude that was introduced in eq. \rf{amp},
\bea
{\cal A}^{\mbox{\tiny SD;A}}_S \!\!\!&=&\!\!\!  \frac{G_{\rm F}}{\sqrt{2}} V_{ud} V_{us} {\rm Re}\,C_{7A}
\times
{\bar{\rm u}} (p_{\ell^-}\!) \gamma_\rho \gamma_5 {\rm v} (p_{\ell^+}\!)
\nonumber\\
&&\!\!\!
\times
\left[ (k+p)^\rho f_+ (s) + (k-p)^\rho f_- (s) \right]  .
\lbl{ampS_SD;A}
\eea
The form factors $f_\pm (s)$ are defined through
(the minus sign is chosen such that the normalization
is $f_+ (0)=1$ in the limit of mass-degenerate $u$, $d$
and $s$ quarks)
\bea
&&\hspace{-0.75cm}
\langle \pi^0 (p) \vert [{\bar s} \gamma_\mu d](0) \vert K_S (k) \rangle =
\nonumber\\
&& = - [ (k+p)_\mu f_+ (s) + (k-p)_\mu f_- (s) ]   .
\lbl{f+_def}
\eea
Having properly defined the form factor ${\cal W}_S (s)$ in terms of
QCD matrix elements, we can now proceed with the evaluation of the 
latter, in the limit $N_c\to\infty$.

For the purpose of this Letter, we will concentrate on the contributions from the four-quark 
operators $Q_1$ and $Q_2$. Indeed, from the results given in Ref. \cite{Buras:1994qa}, one 
infers that the absolute values of the Wilson coefficients $C_{3,4,5,6} (\nu)$  at $\nu = 1$ GeV
are smaller by at least one order of magnitude than the ones of the current-current operators
at the same scale, so that the contributions of the former can be neglected, unless some
of the corresponding matrix elements are enhanced. A more complete analysis \cite{MK_unpub}, including
all the six operators, shows that this is not the case and that in the large-$N_c$ limit
the contribution from the QCD-penguin operators to the form factor ${\cal W}_S (s)$ is indeed marginal. 
Our task then reduces to computing the leading contributions, of order ${\cal O} (N_c)$, to the matrix elements 
$\langle \pi^0 \vert T \{ j_\mu (x) Q_{1,2}(0) \} \vert K_S \rangle$ when $N_c$ becomes large. What 
makes this task possible is the fact that in this limit the four-quark operators factorize 
into the product of two quark bilinears: gluon configurations that would break this factorization
are subleading in the $1/N_c$ expansion. To keep things 
as simple as possible, we only show the expressions obtained when the matrix $\gamma_5$ is handled 
in the 't~Hooft-Veltman scheme \cite{tHooft:1972tcz}. Using naive dimensional regularization \cite{Chanowitz:1979zu} leads to
additional terms in some of the matrix elements \cite{MK_in_prep}, to some extent compensated by the scheme dependence
of the Wilson coefficients. For the operator $Q_1$, one then obtains (from now on, all expressions, unless otherwise specified,
will be understood to hold in the large-$N_c$ limit, and the presence of sub-leading terms in the $1/N_c$ expansion
will not be indicated explicitly)
\begin{eqnarray}
&&\hspace{-0.75cm}
 \langle \pi^0(p)  \vert T \{  j_\rho  (0) Q_1 (x)  \} \vert K^0(k) \rangle =
\nonumber\\
&&\hspace{-0.5cm}
=  - \frac{2}{3} \langle \pi^0(p) \vert [ {\bar s} \gamma_\nu  d ] (0)  \vert K^0(k) \rangle 
\nonumber\\
&&
\times
\langle 0 \vert T \{ [{\bar u} \gamma_\rho u] (x) [{\bar u} \gamma_\nu u ] (0)  \} \vert 0 \rangle
\nonumber\\
&&\!\!\!\!\!
- \, 
\frac{2}{3} \langle 0 \vert [ {\bar s} \gamma_\mu\gamma_5 d ] (0)  \vert K^0(k) \rangle 
\nonumber\\
&&
\times
\langle \pi^0 (p) \vert T \{ [{\bar u} \gamma_\rho u ] (x) [{\bar u} \gamma^\mu (1-\gamma_5) u] (0)  \} \vert 0 \rangle
\nonumber\\
&&\!\!\!\!\!
+ \, 
\frac{1}{3} \langle \pi^0 (p) \vert [ {\bar u} \gamma_\mu\gamma_5 u ] (0)  \vert 0 \rangle 
\\
&&
\times
\langle 0 \vert T \{ [{\bar d} \gamma_\rho d + {\bar s} \gamma_\rho s ] (x) [{\bar s} \gamma^\mu(1-\gamma_5) d] (0)  \} \vert K^0(k) \rangle  .
\nonumber
\end{eqnarray} 
The correlator appearing in the first term on the right-hand side of this expression, 
of the vacuum-polariza\-tion type, is divergent. This divergence
reflects the short-distan\-ce singularity of the time-ordered product \rf{SD_sing} and has to be subtracted in the
${\overline{\rm MS}}$ scheme, as explained previously.
One also immediately notices that $Q_2$ cannot contribute to ${\cal W}_S (s)$ in the large-$N_c$ limit.
The reason for this is easy to understand:
as can be seen from eq. \rf{Q1_Q2}, the operator $Q_2$ factorizes into the product of two 
colour-singlet \textit{charged} currents, $[{\bar s} \gamma_\mu (1-\gamma_5) u]$ and $[{\bar u} \gamma^\mu (1-\gamma_5) d]$, and it is not 
possible to construct non-vanishing matrix elements for these currents with only a neutral pion
and a neutral kaon at disposal. 
After having used invariance under parity, char\-ge conjugation, isospin symmetry, and applied Ward identities \cite{MK_in_prep},
the matrix element of the operator $Q_1$ in the large-$N_c$ limit can be expressed in 
terms of the pion and kaon decay constants $F_\pi$ and $F_K$, respectively, together with: 

\noindent
i) the properly renormalized vacuum-polarization correlation function 
\begin{eqnarray}
&&\hspace{-0.75cm}
i \int d^4 x \, e^{iq\cdot x} \langle 0 \vert T \{ [{\bar u} \gamma_\mu u] (x) [{\bar u} \gamma_\nu u] (0) \} \vert 0 \rangle_{\mbox{\tiny{$\overline{\rm MS}$}}} =
\nonumber\\
&&
= ( q_\mu q_\nu - q^2 \eta_{\mu\nu} ) \Pi_{\mbox{\tiny$\overline{\rm MS}$}} (q^2 ; \nu)    ;
\end{eqnarray}
ii) the form factor $f_+(s)$ already defined in eq. \rf{f+_def};

\noindent
iii) the two vertex functions
\begin{eqnarray} 
&&\hspace{-0.82cm} 
\Gamma_\rho (q,k) =
i \int d^4 x \, e^{i q \cdot x} \langle 0 \vert T \{ [ {\bar d} \gamma_\rho d ] (x)
[ {\bar s} i \gamma_5 d ] (0) \} \vert K^0 (k) \rangle ,
\nonumber\\[-0.2cm]
&&\hspace{-0.8cm}
\\[-0.2cm]
&&\hspace{-0.82cm} 
{\tilde\Gamma}_\rho (q,k) =
i \int d^4 x \, e^{i q \cdot x} \langle 0 \vert T \{ [ {\bar s} \gamma_\rho s ] (x)
[ {\bar s} i \gamma_5 d ] (0) \} \vert K^0 (k) \rangle  .
\nonumber
\end{eqnarray}
These vertex functions each have a kaon pole at $(q-k)^2 = M_K^2$, whose residues involve the kaon form factors
$F_d^{K^0} (q^2)$ and $F_s^{K^0} (q^2)$, defined through the two matrix elements
$\langle K^0 \vert {\bar d} \gamma_\rho d \vert K^0 \rangle$ and $\langle K^0 \vert {\bar s} \gamma_\rho s \vert K^0 \rangle$,
respectively, with normalizations  chosen such that $F_d^{K^0}(0)$ $= -F_s^{K^0}(0) = 1$.
Combined with the Ward identities these vertex functions satisfy, this leads to the convenient representations
($m_s$ stands for the mass of the strange quark, while ${\hat m}$ denotes the common
mass of the up and down quarks in the isospin limit)
\bea  
&&\hspace{-0.75cm}
(m_s + {\hat m}) \Gamma_\rho (q,k) = 
\sqrt{2} F_K M_K^2  \frac{(2k-q)_\rho}{(q-k)^2 - M_K^2} F_d^{K^0} (q^2)
\nonumber\\
&&\hspace{2.0cm}
+ \, \sqrt{2} F_K M_K^2  \frac{F_d^{K^0} (q^2) - 1}{q^2} q_\rho
\nonumber\\
&&\hspace{2.0cm}
+ \,  \sqrt{2} [ q^2 k_\rho - (q\cdot k) q_\rho ] {\cal P} (q^2 , (q-k)^2 )   ,
\nonumber\\
\\
&&\hspace{-0.75cm}
(m_s + {\hat m}) {\tilde\Gamma}_{\rho} (q,k) = 
\sqrt{2} F_K M_K^2  \frac{(2k-q)_\rho}{(q-k)^2 - M_K^2} F_s^{K^0} (q^2)
\nonumber\\
&&\hspace{2.0cm}
+ \, \sqrt{2} F_K M_K^2  \frac{F_s^{K^0} (q^2) + 1}{q^2} q_\rho
\nonumber\\
&&\hspace{2.0cm}
+ \,  \sqrt{2} [ q^2 k_\rho - (q\cdot k) q_\rho ] {\tilde{\cal P}} (q^2 , (q-k)^2 )   .  
\nonumber
\eea
Putting everything together, the expression of the form factor ${\cal W}_S(s)$ at leading-order in the $1/N_c$ expansion reads
\bea
&&\hspace{-0.75cm}
\frac{{\cal W}_S (s)}{16\pi^2 M_K^2} = - \frac{G_F}{\sqrt{2}} V_{us} V_{ud} \bigg\{
f_+ (s) \bigg[ \frac{2}{3} C_1  \Pi_{\mbox{\tiny$\overline{\rm MS}$}} (s ; \nu) 
\nonumber\\
&&\hspace{-0.55cm}
+ \frac{{\rm Re} C_{7V} (\nu)}{4\pi\alpha} \bigg]
+ \frac{2}{3}  C_1 
\bigg[ \frac{ F_\pi F_K M_K^2}{M_K^2-M_\pi^2} \, \frac{F_d^{K^0} (s) + F_s^{K^0} (s)}{s} 
\nonumber\\
&& \hspace{-0.55cm}  
 - \frac{F_\pi}{2} {\cal P} (s , M_\pi^2 ) - \frac{F_\pi}{2} {\tilde{\cal P}} (s , M_\pi^2 ) \bigg] \bigg\}  .
 \lbl{W_S}
\eea

It can be shown that in the large-$N_c$ limit this last expression does not depend on $\nu$ \cite{MK_in_prep}. 
Moreover, the three form factors $f_+(q^2)$, $F_d^{K^0} (q^2)$, $F_s^{K^0}(q^2)$
and the vacuum-polarization function $\Pi_{\mbox{\tiny$\overline{\rm MS}$}} (s ; \nu)$
consist of an infinite number of poles due to zero-width mesonic resonances \cite{tHooft:1973alw} \cite{Witten:1979kh}.
The three form factors  
behave in QCD like $\sim 1/q^2$ for large space-like values of $q^2$. Due to this smooth
asymptotic behavior it is justified to retain only 
the lowest-lying resonance in each case \cite{Peris:1998nj}, i.e. $K^*(892)$ for $f_+$, $\rho(770)/\omega(782)$ for $F_d^{K^0}$, $\phi(1020)$ for $F_s^{K^0}$,
i.e. (we take $M_\omega = M_\rho$)
\bea
&&
f_+ (q^2) = \frac{M_{K^*}^2}{M_{K^*}^2-q^2} , ~
F_s^{K^0} (q^2) = \frac{M_\phi^2}{q^2 - M_\phi^2} ,
\nonumber\\
&&\hspace{-0.25cm}
F_d^{K^0} (q^2) = \frac{M_\rho^2}{M_\rho^2-q^2}   . ~~~
\eea
On the other hand, the function $\Pi_{\mbox{\tiny$\overline{\rm MS}$}} (q^2 ; \nu)$ behaves
as $\sim \ln (-q^2/\nu^2)$ when $q^2\to -\infty$. Clearly, such a logarithmic behavior cannot be 
reproduced by a single resonance pole, and not even by a finite number of such poles, so that
a representation in terms of an infinite number of $J^{PC}=1^{--}$ states cannot be avoided \cite{Witten:1979kh}. 
Fortunately, such representations have been discussed and constructed in 
the literature, see for instance ref. \cite{DAmbrosio:2019xph} and the articles
quoted therein. We will adopt the expression
\begin{eqnarray}
&&
\hspace{-0.75cm}
\Pi_{\mbox{\tiny$\overline{\rm MS}$}} (q^2 ; \nu) = \frac{f_\rho^2 M_\rho^2}{M_\rho^2 - q^2} + \frac{9f_\omega^2 M_\omega^2}{M_\omega^2 - q^2} 
+ \frac{N_c}{12\pi^2} \bigg\{ - \ln (M^2/\nu)
\nonumber\\
&&
\hspace{1.cm}
 + \, \frac{5}{3} -  \psi \left( 3 - \frac{q^2}{M^2} \right) \bigg\}  , ~~~~~
\end{eqnarray}
where $\psi$ denotes the di-gamma function and we have not shown ${\cal O}(\alpha_s N_c)$
corrections, which are known and included in the numerical analysis.
The poles (in the large-$N_c$ limit) due to the $\rho$ and $\omega$ states 
have been shown explicitly. The couplings $f_{\rho,\omega}^2$ 
can be determined from the experimental decay widths $\Gamma(\rho,\omega \to e^+ e^-)$.
For $q^2<0$, $ \psi ( 3 - {q^2}/{M^2})$
is a smooth function, which has a logarithmic asymptotic behavior as $q^2\to -\infty$,
\begin{equation}
\psi \left( 3 - \frac{q^2}{M^2} \right) \sim \ln(-q^2/M^2) - \frac{5}{2} \frac{M^2}{q^2} + {\cal O} (M^4/q^4)
\end{equation}
thus reproducing the leading perturbative expression of $\Pi_{\mbox{\tiny$\overline{\rm MS}$}} (q^2 ; \nu)$.
For $q^2>0$ the di-gamma function sums a series of equidistant poles located at the values $q^2=M_n^2\equiv (n+2)M^2$,
\begin{equation}
\psi \left( 3 - \frac{q^2}{M^2} \right) = - \gamma_E + \frac{3}{2} + \sum_{n\ge 1} \frac{1}{n+2} \frac{q^2~~}{q^2-M_n^2}  ,
\end{equation}
where $\gamma_E$ is the Euler constant.
We still need to fix the value of the mass scale $M$. This can be done upon using the following 
constraint on the Adler function ${\cal A} (q^2) \equiv - q^2 (\partial\Pi(q^2)/\partial q^2)$: for large Euclidian 
values of the momentum $q$, the behavior of ${\cal A}(q^2)$ in QCD cannot display a term  $\sim1/q^2$ in the chiral limit \cite{Peris:1998nj}.
Neglecting ${\cal O}(\alpha_s N_c)$ corrections, this condition requires (we have taken $f_\omega M_\omega \sim f_\rho M_\rho/3$, as required in the combined
large-$N_c$ and isospin limits and as also reproduced by data)
\be
M^2 = \frac{16 \pi^2}{5} \frac{3}{N_c} \, f_\rho^2 M_\rho^2   .
\ee
For $N_c=3$ and $f_\rho M_\rho \sim 154$ MeV, this yields $M \sim  0.87$ GeV and $M_1 \sim 1.5~{\rm GeV}$, which is quite 
reasonable, this last value being comparable to the mass of the $\rho(1450)$, the first $J^{PC}=1^{--}$ resonance after the 
$\rho(770)$ \cite{ParticleDataGroup:2022pth}.

It remains to discuss the functions ${\cal P} (q^2 , (q-k)^2 )$ and  ${\tilde{\cal P}} (q^2 , (q-k)^2 )$. 
These two functions account for the poles produced by zero-width radial excitations of the kaon, i.e. $K'$, $K''$,.... The
first of these states can, for instance, be identified with the $K(1460)$ resonance
in the real world where $N_c=3$. 
Two important observations concerning them can be made and exploited \cite{MK_unpub}. 
First, the behavior of $\Gamma_\rho (q,k)$ and  ${\tilde\Gamma}_\rho (q,k)$ 
at large space-like values of $q^2$, as determined by the operator-product expansion, shows that the leading short-distance term is saturated by the contribution 
due to their longitudinal parts, i.e. the kaon poles. Therefore, the functions ${\cal P} (q^2 , (q-k)^2 )$ and  ${\tilde{\cal P}} (q^2 , (q-k)^2 )$
provide only subdominant contributions at short distances. Second, the poles due to the radial excitations of the kaon will come with
the factors of the kaon poles replaced by $F_{K'} M_{K'}^2/[((q-k)^2 - M_{K'}^2]$, where $F_{K'}$ is the decay 
constant of the radial excitation $K'$ of mass $M_{K'}$. Since these states do not become Goldstone bosons in 
the chiral limit, $F_{K'}$ must vanish linearly with vanishing quark masses. Indeed, estimates based on QCD sum rules 
\cite{Narison:1982gw} \cite{Maltman:2001gc}
give values much smaller than the kaon decay constant $F_K$ for the first of these radial excitation, 
e.g. $F_{K'} = 21.4(2.8)~{\rm MeV}$ \cite{Maltman:2001gc}, and an even smaller value for the second radial excitation.
In addition, the factor $M_{K'}^2$ in the residue of the pole in cancelled 
by the denominator when one eventually takes $q=k-p$, so that  $F_{K'} M_{K'}^2/[((q-k)^2 - M_{K'}^2]$ becomes
$\sim - F_{K'}$. Barring any large enhancement due to the electromagnetic transition form factors
$F^{K^0 K'}_{u,s} (q^2)$ that replace $F_{u,s}^{K^0} (q^2)$, this indicates that the 
contributions of ${\cal P} (q^2 , (q-k)^2 )$ and  ${\tilde{\cal P}} (q^2 , (q-k)^2 )$ 
are highly suppressed as compared to the contributions from the kaon poles, 
which leads us to make the approximations 
${\cal P} (q^2 , (q-k)^2 ) \simeq 0 , ~ {\tilde{\cal P}} (q^2 , (q-k)^2 ) \simeq 0$.

We have now all the elements at our disposal in order to answer the three questions
listed at the beginning of this Letter. We use the values of the Wilson coefficients
at the scale $\nu = 1~{\rm GeV}$
given in ref. \cite{Buras:1994qa}, and the values of the remaining quantities 
are taken from ref. \cite{ParticleDataGroup:2022pth}.
The values shown below result from the average of those obtained with
the 't Hooft-Veltman scheme and with the naive dimensional regularization scheme.

\noindent
$\quad\bullet$~The predictions for the branching ratios read
\bea
&&
{\rm Br} (K_S \to \pi^0 e^+ e^-)\vert_{ m_{ee} > 165~{\rm MeV}} = 2.9(1.0) \cdot 10^{-9}
\nonumber\\
&&
{\rm Br} (K_S \to \pi^0 e^+ e^-) = 5.1(1.7) \cdot 10^{-9}  , 
\nonumber\\
&&
{\rm Br} (K_S \to \pi^0 \mu^+ \mu^-) = 1.3(0.4) \cdot 10^{-9},
\lbl{Br_results}
\eea
where a conservative relative uncertainty of ${\cal O}(1/N_c) \sim 30\%$, accounting 
for sub-leading effects in the $1/N_c$ expansion, has been applied with $N_c=3$.
The first value agrees well with the measurement by the NA48/1 experiment \cite{NA481:2003cfm}
(the first error is statistics, the second systematics)
\be
{\rm Br} (K_S \to \pi^0 e^+ e^-)\vert_{ m_{ee} > 165~{\rm MeV}} = (3.0^{+1.5}_{-1.2} \pm 0.2) \cdot 10^{-9}  .
\ee
When extrapolated to the full range of the di-lepton invariant mass $m_{ee}$ 
with a form factor equal to unity (i.e. with $a_S=1$, $b_S=0$ and no pion loop) the total
branching fraction is quoted as ${\rm Br} (K_S \to \pi^0 e^+ e^-) = (5.8^{+2.8}_{-2.3} \pm 0.8 ) \cdot 10^{-9}$ \cite{NA481:2003cfm},
which also agrees rather well with the value in eq. \rf{Br_results}.
The agreement is less good in the case of the decay into a muon pair, where the experimental value obtained
by the NA48/1 collaboration is given as \cite{NA481:2004nbc}
${\rm Br} (K_S \to \pi^0 \mu^+ \mu^-) = (2.9^{+1.5}_{-1.2}{\rm (stat)} \pm 0.2{\rm (syst)}) \cdot 10^{-9}$, but the 
uncertainties are still large. Finally, we also mention that within the range $s\in[0,M_K^2]$, which 
covers the  phase space of the $K_{S,L} \to \pi^0 \ell^+ \ell^-$ decays, the form  
factor \rf{W_S} is well described by the quadratic polynomial
\be
{\cal W}_S (s) \sim G_{\rm F} [ 0.92 M_K^2 + 0.64 s + 0.39 s^2/M_K^2]  .
\lbl{WS_quad}
\ee

\noindent
$\quad\bullet$~The function ${\cal W}_S (s)$ being positive, cf. eq. \rf{WS_quad},
the coefficients $C_{\rm int}^{(\ell)}$ in eq. \rf{Br_KL} are also positive. Numerically, we obtain
\be
C_{\rm int}^{(e)} = + 7.8(2.6) \, \frac{y_{7V}}{\alpha}
,~ C_{\rm int}^{(\mu)} = + 1.9(0.6) \, \frac{y_{7V}}{\alpha},
\ee
where we have written ($y_{7V}$ is positive \cite{Buras:1994qa})
\be
V_{ud} V_{us} {\rm Im} \, C_{7V} (\nu=1~{\rm GeV}) = - ({\rm Im} \, \lambda_t) y_{7V}  .
\ee
The interference between direct and indirect CP violation
in the branching ratio for $K_L\to\pi^0\ell^+\ell^-$ is therefore unambiguously predicted 
to be constructive in the large-$N_c$ limit of QCD. 

\noindent
$\quad\bullet$~The amplitude of the CP-violating transition $K_2^0\to\pi^0\gamma^*\to \pi^0\ell^+\ell^-$ has the same 
structure as given in eq. \rf{amp}, provided one makes the replacements
$C_1 \to 0$, ${\rm Re}\,C_{7X} \to i{\rm Im}\,C_{7X}$, $X=V,A$,
in eqs. \rf{ampS_SD;A} and \rf{W_S}.
In addition, as already mentioned, the matrix elements of the QCD penguin
operators show no particular enhancement as compared to the matrix element of $Q_1$ \cite{MK_unpub},
while the imaginary parts of their Wilson coefficients at the scale $\nu=1~{\rm GeV}$ 
are about one order of magnitude smaller (in absolute value) 
than $\vert{\rm Im}\,C_{7X}\vert/\alpha$ at the same scale \cite{Buras:1994qa}. 
The approximation consisting in keeping only the contribution from the Gilman-Wise operators
is therefore also supported by the large-$N_c$ limit of QCD.

Before concluding, let us briefly discuss the case of ${\cal W}_+ (s)$ in the context of the large-$N_c$ limit.
The main difference with ${\cal W}_S (s)$ lies in the fact that now the operator $Q_2$ will 
contribute. Actually, the contribution of the operator $Q_1$ to ${\cal W}_+ (s)$ is now
limited to the term proportional to $\Pi_{\mbox{\tiny$\overline{\rm MS}$}} (s ; \nu)$ 
in eq. \rf{W_S}, whereas expressions similar to the remaining terms in this equation will instead 
be produced by $Q_2$. Since $C_1 (\nu) \simeq - C_2 (\nu)/2$ at $\nu \simeq 1~{\rm GeV}$, this leads to an almost complete 
numerical cancellation between the two contributions \cite{MK_in_prep}, leaving only the small contributions from 
the QCD penguin operators as a remainder. The almost vanishing values of $a_+$ and $b_+$ predicted 
by the large-$N_c$ limit do thus not at all account for the measured values \cite{NA62:2022qes}, and sub-leading terms 
in the $1/N_c$ expansion must become important in this case. This is quite in line with
the result of ref. \cite{DAmbrosio:2018ytt}, where a crude unsubtracted dispersive evaluation of the 
contribution from two-pion states to ${\cal W}_+ (s)$, suppressed in the large-$N_c$ limit but this time 
enhanced by the $\Delta I = 1/2$ rule, produced values of $a_+$ and $b_+$ already reasonably close to the experimental ones.

To summarize, we have outlined the computation of the amplitudes for the kaon decay 
modes $K_{S,L}\to\pi^0\gamma^*\to\pi^0\ell^+\ell^-$ in the large-$N_c$ limit of QCD. We 
have shown that this framework is predictive as it allows to answer a few questions of phenomenological 
relevance for the possibility to experimentally probe the standard-model's
flavour structure at short distances. A more detailed account of the calculation 
and further implications will be provided in ref. \cite{MK_in_prep}.
For completeness, let us also mention that the proposal \cite{Isidori:2005tv} \cite{Christ:2015aha} \cite{Christ:2016mmq}
to investigate the $K\to\pi\ell^+\ell^-$ decay modes in the framework 
of lattice QCD is being actively pursued by the RBC and UKQCD collaborations.
A first result for ${\cal W}_+ (s)$ at $s/M_K^2 = 0.013(2)$ with physical values of the pion and kaon masses
was published recently \cite{RBC:2022ddw}. It corresponds
to only a single lattice spacing and still shows quite large uncertainties due to the difficulty
of extracting the signal from the statistical noise. Substantial improvements are however expected 
during the next decade for this and for other rare kaon decay modes \cite{Blum:2022wsz} \cite{Anzivino:2023bhp}.
In the meantime, the quest for a better theoretical and phenomenological understanding
of rare kaon decay modes is certainly worth being pursued as well. The large-$N_c$ limit may shed light
on other processes than the ones studied here and bring to the fore interesting dynamical aspects
and/or quantitative information. 
Of course, cancellations can also happen in other amplitudes than the one for $K^\pm\to\pi^\pm\ell^+\ell^-$,
but most probably only  a case-by-case study can eventually reveal 
which observables are actually affected or not.  

\indent

\noindent
\textbf{Acknoledgements}: One uf us (MK) would like to thank the Department of Physics 
of the Universit\`a degli Studi Federico II di Napoli for the warm 
hospitality extended to him, and the Sezione di Napoli of INFN for financial support.

\end{document}